\documentclass[traditabstract]{aa-sarah}  
\usepackage{graphicx}
\usepackage{txfonts}
\usepackage{natbib}
\bibpunct{(}{)}{;}{a}{}{,} 
\usepackage[normalem]{ulem} 
\usepackage{color} 
\begin{document}
\title{Red giants in the outer halo of the elliptical galaxy NGC 5128 / Centaurus A}

\author{Sarah~A.~Bird
  \inst{\ref{inst1}}
\thanks{E-mail: sarah.bird@utu.fi}
  \and
  Chris~Flynn
  \inst{\ref{inst2},\ref{inst4}}
  \and
  William~E.~Harris
  \inst{\ref{inst3}}
  \and
  Mauri~Valtonen
  \inst{\ref{inst2}}
}

\institute{Tuorla Observatory, Department of Physics and Astronomy, 
  University of Turku, V\"ais\"al\"antie 20, FI-21500 Piikki\"o, Finland
  \label{inst1}\\
  \and
  FINCA - Finnish Centre for Astronomy with ESO, 
  University of Turku, V\"ais\"al\"antie 20, 
  FI-21500 Piikki\"o, Finland\label{inst2}\\
  \and
  Centre for Astrophysics and Supercomputing, 
  Swinburne University of Technology, 
  Hawthorn VIC, 3122, Australia\label{inst4}\\
  \and
  Department of Physics and Astronomy, McMaster University, 
  Hamilton ON, L8S 4M1, Canada\label{inst3}\\
}

\date{Received 14 January 2014 / Accepted 14 October 2014}


\abstract{We used VIMOS on VLT to perform $V$ and $I$ band imaging of the outermost halo of \object{NGC 5128}~/~\object{Centaurus A} ($(m-M)_0=27.91\pm0.08$), 65~kpc from the galaxy's center and along the major axis. The stellar population has been resolved to $I_0 \approx 27$ with a 50\% completeness limit of $I_0 = 24.7$, well below the tip of the red-giant branch (TRGB), which is seen at $I_0 \approx 23.9$. The surface density of NGC 5128 halo stars in our fields was sufficiently low that dim, unresolved background galaxies were a major contaminant in the source counts. We isolated a clean sample of red-giant-branch (RGB) stars extending to $\approx 0.8$~mag below the TRGB through conservative magnitude and color cuts, to remove the (predominantly blue) unresolved background galaxies. We derived stellar metallicities from colors of the stars via isochrones and measured the density falloff of the halo as a function of metallicity by combining our observations with HST imaging taken of NGC 5128 halo fields closer to the galaxy center. We found both metal-rich and metal-poor stellar populations and found that the falloff of the two follows the same de Vaucouleurs' law profiles from $\approx 8 $ kpc out to $\approx$ 70 kpc. The metallicity distribution function (MDF) and the density falloff agree with the results of two recent studies of similar outermost halo fields in NGC 5128. We found no evidence of a ``transition'' in the radial profile of the halo, in which the metal-rich halo density would drop rapidly, leaving the underlying metal-poor halo to dominate by default out to greater radial extent, as has been seen in the outer halo of two other large galaxies. If NGC 5128 has such a transition, it must lie at larger galactocentric distances.}


\maketitle
%

\section{Introduction}

The spatial distribution of stars of differing metallicities within a galaxy
gives clues to the process of its formation and evolution. The very earliest
stars in giant galaxies -- the most metal-poor halo stars and globular clusters
-- may have formed before the onset of hierarchical merging, within small
pregalactic dwarfs that populated the large-scale dark-matter potential well
\citep{Bullock2005,Naab2009,Oser2010}. Today, these relic stars would be found
in giant galaxies in a sparse and extremely extended ``outermost halo''
component \citep{Harris2007}.

Few have approached the extraordinarily difficult task of finding
  clear traces of this component in large galaxies and of deconvolving it from
  the more obvious and metal-rich spheroid component generated either later by
  major mergers or earlier through dissipative collapse. An effective way to
isolate the halo stellar population in nearby galaxies and derive its
MDF is through direct multicolor photometry
of its resolved stars, as opposed to indirect methods from integrated
light. For nearby galaxies, to distances $\le20$~Mpc, resolved-star photometry
is readily tractable either with the Hubble Space Telescope (HST) or with 8
m class ground-based telescopes.

Surprising evidence has begun to be uncovered that the metal-rich stellar
density falls off more rapidly in the outermost halo than does the metal-poor
distribution, with a ``transition radius'' around $10-12~R_\mathrm{eff}$ (where
$R_\mathrm{eff}$ is the effective radius within which half of the visible light
is located). Evidence of such a transition has been found in two very
different types of galaxies, the SA(s)b spiral galaxy \object{NGC 224}
(\object{M31}) (absolute visual magnitude $M_V=-21.78$ mag)
\citep{Kalirai2006,de_Vaucouleurs1991} and the E1 elliptical \object{NGC 3379}
($M_V=-20.82$ mag) \citep{Harris2007,de_Vaucouleurs1991}. If this transition is
a feature of all major galaxies, then the metal-poor stars could be readily
probed beyond $10~R_\mathrm{eff}$ without the interference of metal-rich stars,
which otherwise dominate throughout the central bulge and inner halo. Such a
transition feature has not been studied thoroughly, for one reason, because
observations are needed at large distances from the galactic
center. 
\citet{Harris2007-ngc3377} study the halo of \object{NGC 3377} and find no transition, but their field reaches to less than
  $4~R_\mathrm{eff}$, so the $10-12~R_\mathrm{eff}$ region remains entirely
  unprobed.

In this study we search for a similar transition to a metal-poor halo within
another galaxy type, the E / S0 galaxy \object{NGC 5128} / \object{Centaurus A}
\citep{Harris1999} ($M_V=-21.39$ mag \citep{de_Vaucouleurs1991} and
$R_\mathrm{eff}=5.6$~kpc or $305^{\prime\prime}$ \citep{Dufour1979}). Previous
studies of NGC 5128 have used the HST cameras to resolve individual RGB stars in fields at galactocentric distances from 8 to 40
kpc, equivalent to 2 to 8 $R_\mathrm{eff}$
\citep{Soria1996,Harris1999,Marleau2000,Harris2000,Harris2002,Rejkuba2005,Gregg2004}.
Isochrones were fit to the color-magnitude diagram (CMD), and the MDF was then
derived by interpolation between the RGB isochrones.  To date, all these fields
-- which cover the outer bulge and inner- to mid-halo regions -- have revealed
a broad spread of metallicity with the majority of stars having [M/H] $> -1$
and peaking at [M/H]$\sim -0.5$. If a transition to a predominantly
metal-poor halo exists and is at a similar distance to what was found in M31
and NGC 3379, then it should occur beyond a projected galactocentric distance
$R_\mathrm{gc} \simeq 60$~kpc, assuming $R_\mathrm{eff} = 5.6$ kpc
\citep{Dufour1979} for NGC 5128.

\object{NGC 5128} is by far the closest easily observable giant elliptical, at
a distance of $d = 3.8$ Mpc \citep{Harris2010}. It is the centrally dominant
galaxy in the Centaurus group, with $\approx$ 30 neighboring
  galaxies at projected radii of $0.05 - 1.0$ Mpc from NGC 5128
  \citep{Karachentsev2005}. The galaxy itself extends over 2$^\circ$ on the
sky with 1$^{\prime}$ corresponding to approximately 1 kpc at the distance of
the galaxy. Its 27 neighboring galaxies, in the Centaurus group, extend over
$25^\circ$ on the sky \citep{Karachentsev2005}. Clear signs of a merger or
accretion history show up in the faint arcs and shells
\citep{Malin1983,Peng2002} extending out to $r\approx 20$ kpc and are also
indicated by the presence of the inner rotating ring of dust and gas with a
radius of $r\approx 3$~kpc \citep{Graham1979,Peng2002}.  Planetary nebulae and
globular clusters have been found as far as 80 kpc and 40 kpc
\citep{Peng2004,Peng2004a,Woodley2010AJ}, respectively, from the galaxy's
center. Recent analysis of the stars in the 40~kpc field \citep{Rejkuba2005}
and the globular clusters \citep{Beasley2008,Woodley2010ApJ} show that their
mean age is in the range $9-12$ Gyr, demonstrating that these fields contain
ancient stars. \citet{Woodley2007,Woodley2010AJ} estimate the dynamical mass as
$1.0\times10^{12}M_\odot$ out to a radius of $r=1.3\degr$ (or 90~kpc) from
analysis of the kinematics of halo planetary nebulae, globular clusters, and
Centaurus-group satellites.

As we were completing this work, two studies of particular interest were published. The first is a similar study to ours by \citet{Crnojevic2013} also of
fields in the outer halo of NGC 5128 and also taken in
the $V$ and $I$ bands with VLT/VIMOS. They observed two fields each along both
the major and minor axes covering $30-85$~kpc, corresponding to $\approx5-14~R_\mathrm{eff}$. Second, \citet{Rejkuba2014} used HST ACS and WFC3 to image NGC 5128 reaching even more distant fields: 60, 90, and 140 kpc ($\approx11$, 16, and $25~R_\mathrm{eff}$) along the major axis and 40 and 90 kpc ($\approx7$ and $16~R_\mathrm{eff}$) along the minor axis. Our results are in good agreement and are discussed in Sec. \ref{mdf}.

We now summarize the organization of our paper. In Sec. \ref{Observations} we
discuss our outer halo observations of NGC 5128 and in Sec. \ref{Reduction} we
describe the data reduction and completeness of our data. Contamination of the
RGB by foreground stars and unresolved background galaxies is carefully
evaluated in Sec. \ref{cmd}, and a clean sample of RGB stars in NGC 5128's
outer halo is derived. The CMD of our RGB stars is used to derive the MDF of
our stars in Secs. \ref{cmd} and \ref{mdf}, although our sample is restricted
to stars with [M/H] $>-1$ because of the cuts which remove background
galaxies. In Sec. \ref{mdf}, we discuss our findings of a
  dominating metal-rich stellar halo and the underlying metal-poor halo. We
  measure the density falloff of our metal-rich and metal-poor stars, both of
  which follow the same de Vaucouleurs' profile, and we find no evidence of a
  transition from a metal-rich to a metal-poor halo despite probing to a
  distance of $\approx 70$~kpc. Any such transition must lie further out
  still. The summary and our conclusions are found in Sec. \ref{Conclusions}.


\section{Observations}
\label{Observations}

\begin{figure}
  \centering 
  \includegraphics[width=10cm]{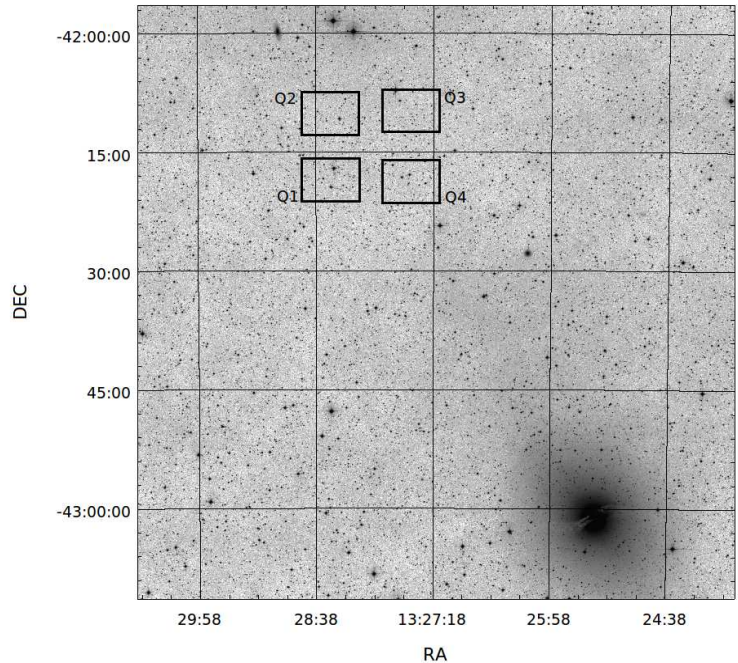}
  \caption{Location and relative size of our VIMOS halo field (seen upper
    left), shown with respect to the center of \object{NGC 5128} /
    \object{Centaurus A} (lower right). The four boxes mark the regions covered
    by each of the four VIMOS CCDs, and lie along the major axis. The image is
    $70^{\prime}\times70^{\prime}$ and is from the STScI Digital Sky
    Survey. North is up and east is to the left.}
  \label{cenA_quad}
\end{figure}

We have imaged an outer halo field of NGC 5128 in $V$ and $I$ bands, during ESO
Periods 83 and 87 (June - July 2009, Run ID 083.B-0036A, and April - July 2011,
Run ID 087.B-0135A) using service mode at the VLT. The camera was VIMOS, which
has a field of view of $4\times (7' \times 8')$ (i.e. 224 square arcminute),
at an image scale of $0.205^{\prime\prime} /$~pixel.

Our field is centered at $\alpha = 13^{\mathrm{h}} 27^{\mathrm{m}}
59^{\mathrm{s}}$ and $\delta=-42^{\circ} 14^{\prime} 50^{\prime\prime}$
(J2000.0) and lies at a projected distance of $59.9^{\prime}$ northeast of the
galaxy's center along the isophotal major axis (see Fig.~\ref{cenA_quad}).  The
linear projected distance to the center of our field is $\approx 66$ kpc, or 12
$R_\mathrm{eff}$, where we have adopted
$R_\mathrm{eff}=5.6$ kpc or $305^{\prime\prime}$ \citep{Dufour1979}.

Pre-observation simulations indicated that the RGB stars would be resolved to
at least a magnitude below the TRGB at $I_\mathrm{TRGB}=24.1$
\citep{Harris2010} in reasonable exposure times for seeing of 0.6 arcsec or
better. In total, we obtained $8\times705$ sec in $I$ and $19\times965$ sec in
$V$, plus one additional 88 sec exposure in $V$; but about half of all the
acquired data were classified by ESO as ``out of spec'' (seeing was worse than
0.6 arcsec or had failed due to technical reasons).  The four $I$ band frames
taken during Period 87 and all 20 $V$ band frames from both periods were,
despite some of which being ``out of spec,'' of sufficiently high quality for
analysis, and were chosen to produce combined images in each band. We did not
use the four $I$ band images from Period 83 because of strong
fringing. The median seeing was 0.54 arcsec and 0.68 arcsec for
  the $I$ and $V$ band frames used for analysis, respectively.  The total
yield in exposure times used was 47 minutes in $I$ band (4 frames with
$\mathrm{FWHM}=0.5^{\prime\prime}$ combined) and 5.1 hours in $V$ (20 frames
with $\mathrm{FWHM}=0.7\arcsec$ combined).

\section{Data Reduction and Completeness}
\label{Reduction}

\subsection{Image Alignment}

\begin{figure}
\centering
\includegraphics[width=8cm]{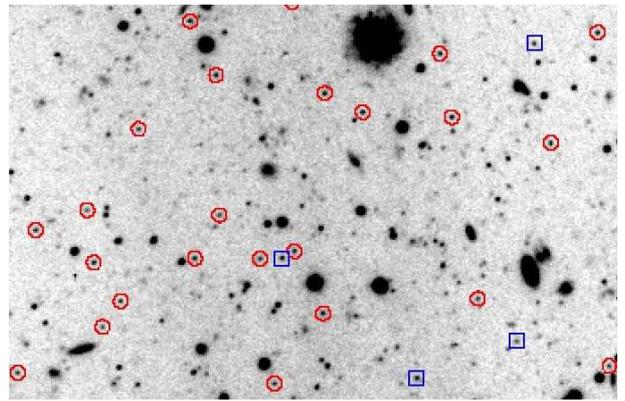}
\caption{Small ($1^\prime.6 \times 0^\prime.7$) section in $I$ band of the
  NGC~5128 / Centaurus A halo, excised from the southeast corner of VIMOS
  Quadrant 4, and located at $\alpha
  =13^{\mathrm{h}}27^{\mathrm{m}}41^{\mathrm{s}}$ and
  $\delta=-42^{\circ}~20^{\prime}~41^{\prime\prime}$. This is our final
  combined image produced from 47 minutes of observations. The
    red circles and blue squares mark point sources classified as metal-rich
    and metal-poor RGB stars respectively (see Sec. \ref{mdf}), with
  $23.9<I_0<24.7$, within the $50\%$ completeness selection criteria detailed
  in Sec. \ref{mock}. [See the electronic edition of the Journal for a color
    version of this figure.]  }
\label{I_Q4c_rg}
\end{figure}

Pointing differences on the sky resulted in shifts between our images of up to
15 pixels. As the image plane in this wide field of view camera has significant
non-linear distortions on these scales, we required $15-20$ bright, isolated,
unsaturated stars in each of the four quadrants of the preprocessed images
supplied by ESO for image alignment. Once the $15-20$ reference stars were
found and matched in all exposures, the IRAF programs GEOMAP, GEOTRAN, and
IMCOMBINE were used to align and combine the images. IMCOMBINE images of each
quadrant were made for the two bands $I$ and $V$, making a total of 8
images. The combined image was cleaned of cosmic rays, bad pixels, and other
artifacts. Bad edges were present along the CCDs, and these were trimmed from
the final combined frames so that all have the same size, leaving $\approx70\%$
of the field of view, or $4\times 6.9'\times 5.5'$ (corresponding to $185~
\mathrm{kpc}^{2}$ at the source). A small segment of Quadrant 4 in the $I$ band
is shown in Fig. \ref{I_Q4c_rg}: we have marked the metal-rich
  and metal-poor point sources (see Sec. \ref{mdf}) in the field by red
circles and blue squares. These point sources have magnitudes of
$23.9<I_0<24.7$ and are at least $50\%$ complete (see Sec. \ref{mock}).

\subsection{Photometric Calibration}

Exposures of 2 seconds duration of Landolt standard star fields
\citep{Landolt1992,Stetson2000} were provided by ESO for most nights our data
were taken. Typically many dozens of Landolt standards could be found in these
frames, and we used these to set the photometric zeropoints of the camera in
each band. The individual frames of the combined NGC 5128 image were observed
at airmasses $z$ which range from $z = 1.05$ to 1.19 in $I$ and $z = 1.05$ to
1.42 in $V$ and we selected standard frames taken at similar mean airmass as
the NGC 5128 fields, so that uncertainties due to the airmass corrections in
$I$ and $V$ are much less than $0.01$ mag and $\approx 0.02$ mag
respectively. This is negligible compared to the dominant source of error,
which is Poisson noise.

The average scatter of the standard star magnitudes, after fitting for a small
color term in the zeropoint relations, was $\pm$ 0.03 for $I$ band and $\pm$
0.02 for $V$ band, for standards with $V_0 > 21$ (stars brighter than $I_0=21$
are saturated), and we are thus confident of our zeropoints to $<0.01$~mag. We
calculated the average aperture correction to large radius for $V$ and $I$ as
0.34 and 0.19 magnitudes, respectively, and, for the
  \citet{Cardelli1989} reddening law of $A_V/E(V-I)=2.571$ ($A_V/E(B-V)=3.1$),
  we used $E(V-I)=0.14$ for the reddening and $A_I=0.22$ for the extinction
  corrections \citep{Schlegel1998}. We use the \citet{Schlegel1998} dust maps through the Galactic Dust Reddening and Extinction Service provided by NASA/IPAC Infrared Science Archive\footnote{Extinction Service available at http://irsa.ipac.caltech.edu/applications/DUST/} to estimate the variation of extinction over our field
  as less than 0.02 mag, negligible compared to other sources of error.
  
\subsection{Source Classification}
\label{SubSecSharp}
Point sources in the field were found using two passes through the DAOPHOT II
\citep{Stetson1992} sequence FIND/PHOT/ALLSTAR \citep{Stetson1987,
  Stetson1994}. On average, 9 bright, isolated stars for $I$ and 34 stars for
$V$ band were used by DAOPHOT in each quadrant with which to measure the point
spread function (PSF). We used a detection threshold of 1.5 standard deviations
above sky noise to produce source catalogs. After DAOPHOT located objects in
both $I$ and $V$ bands, the two lists of objects were matched, via our own
software, which pairs stars together of different bands using an initial
estimate of the pixel shift between the two frames, and iterating to get the
final alignment. The images reach to $I_0 \approx 27.0$ and $V_0 \approx 28.4$, or
roughly three magnitudes deeper than the TRGB at
$I_0=23.9$ \citep[we have corrected this value for extinction]{Harris2010}.

We used a similar ``sharpness'' parameter to the one in DAOPHOT's routine
ALLSTAR to measure how point-like the sources are. Specifically, we measured the
ratio of the flux in the profile of objects between 2.5 to 3.5 pixels from the
center, compared to the flux in the central pixel (note well that the data are
modestly oversampled, as the pixel size is 0.2 arcsec, compared to the seeing
disk of 0.6 arcsec). Our sharpness parameter for an object is the
difference between this ratio and the same ratio for the PSF. Sharpness values
were measured for the $I$ band images only, as they are much sharper than our
$V$ band images (the seeing was better in $I$, and far fewer images have been
combined, to make the final $I$ image). 

\begin{figure}
\centering 
\includegraphics[width=8cm]{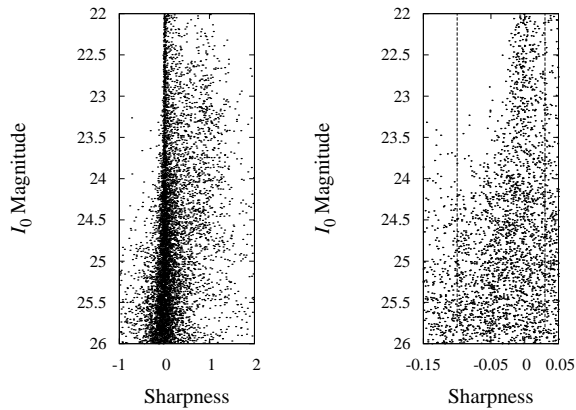}
\caption{$I$-band-derived sharpness parameter versus $I_0$~band magnitudes of objects found in
  both $I$ and $V$~band combined images of VIMOS Quadrant 2 (objects detected in only one filter have been excluded). In the left panel,
  stars lie along the ridge of objects at the sharpness value of zero, while
  non-stellar or double stars lie mainly in the spread of points to the right
  of the ridge. The right panel is a zoomed in version of the left panel. The
  dashed lines mark the sharpness values of $-0.10<\mathrm{sharpness}<0.03$, in
  which range we have defined objects as point sources.  }
\label{sharp}
\end{figure}

In Fig.~\ref{sharp} (left panel) we show the distribution of the 
$I$-band-derived sharpness parameter for objects found in both the combined VIMOS (Quadrant 2) $I$ and $V$ band images (objects found through only one filter have been excluded). In the magnitude range $23<I_0<25$, careful
inspection by eye showed that objects with sharpness $>0.05$ were clearly
extended (see Fig.~\ref{sharp}), while objects $<-1.0$ were predominantly
artifacts, such as seen around overexposed stars. We conservatively defined
point sources as having $-0.10<\mathrm{sharpness}<+0.03$ in our $I$~band image
(i.e. between the dashed lines shown on the right panel of Fig.~\ref{sharp}),
after doing tests with mock stars inserted into the frames, as described in the
next section. We found a total of 12,237 point sources covering a range of magnitude $21 < I_0 < 27$ and color $-1 < (V-I)_0 < 5$ (See the CMDs in Fig \ref{colormag_quad}$-$\ref{colormag}).

\subsection{Mock Stars and Sample Completeness Limits}
\label{mock}

\begin{figure}
\centering 
\includegraphics[width=8cm]{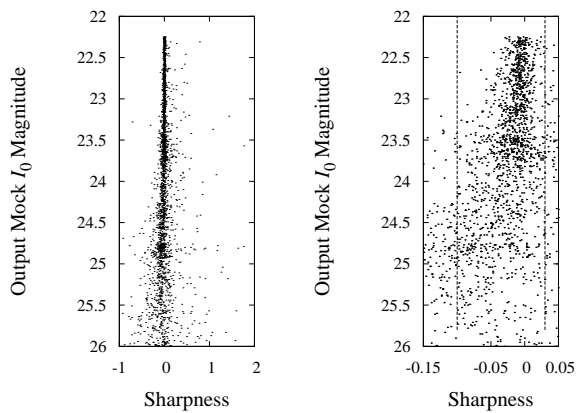}
\caption{$I$-band-derived sharpness versus $I_0$~magnitudes of the mock stars recovered in both
  $I$ and $V$ filters (see also Fig. \ref{mockhistVI}) as described in
  Sec. \ref{mock}. The left panel shows mock star sharpnesses against $I_0$~band
  magnitude, and the right panel a zoomed version of the same plane. The great
  majority of the mock stars in the magnitude range $24 < I_0 < 25$ lie in the
  region $-0.10<\mathrm{sharpness}<0.03$, which we use to select our point
  sources. 
}
\label{mocksharp}
\end{figure}

\begin{figure}
\centering 
\includegraphics[width=8cm]{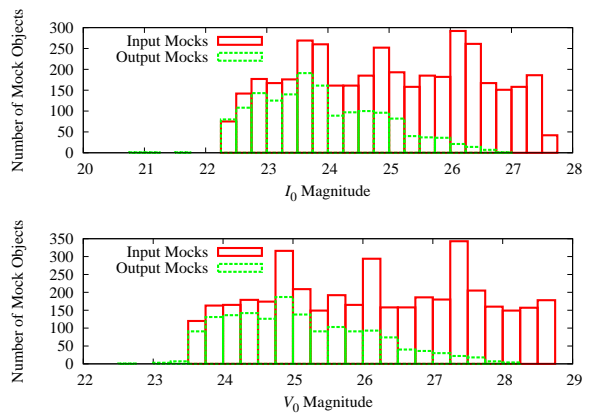}
\caption{Histograms of the number of mock stars in 0.25 magnitude wide bins of
  $I_0$ and $V_0$. In the top panel, the solid red lined histogram includes all
  4000 mock $I_0$ band stars randomly placed in VIMOS Quadrant 1 (see the lower
  panel for the corresponding $V$~band histogram). The mock stars are given
  random $I_0$~magnitudes in the range $22<I_0<28$. In the top panel, the green dot
  dashed histogram displays the 1571 mock stars found and measured by DAOPHOT,
  matched with those found in the $V$ band image, and having
  $-0.1<\mathrm{sharpness}<0.03$ as measured from the $I$ band image (see the
  lower panel for the corresponding histogram of the 1571 objects measured in
  the $V$ band image). 
[See the electronic
edition of the Journal for a color version of this figure.]
}
\label{mockhistVI}
\end{figure}

\begin{figure}
\centering 
\includegraphics[width=8cm]{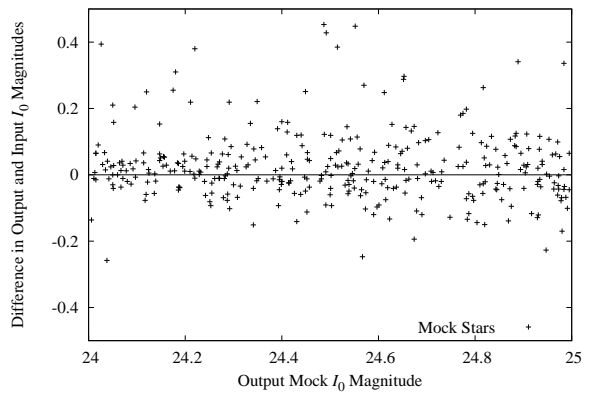}
\caption{$I_0$ magnitudes for mock stars recovered and
  matched from the 4000 mock stars randomly placed within $I$ and $V$ band
  images of VIMOS Quadrant 1 (see Fig.~\ref{mockhistVI} and
  Sec. \ref{mock}). 
  }
\label{IinVsIout}
\end{figure}

We have estimated completeness limits for our source lists and verified our
selection criteria for point sources by placing mock stars into the final
aligned images in each band and re-running DAOPHOT (FIND/PHOT/ALLSTAR), to
measure the recovery rate for the new sources and estimate any offset between
their input and output magnitudes. We inserted 4000 mock stars into the $I$ and
$V$~band VIMOS Quadrant 1 field, using the PSF measured for this quadrant from
bright sources. 
The stars were assigned magnitudes randomly in the range
$22<I_0<28$ and a color of $(V-I)_0 = 1.2$, similar to the color of the RGB
stars in our field. 
We recovered 1571 of these mock stars, which both are matched in $V$ and $I$ as well as fall within our defined sharpness range for point sources of $-0.10<\mathrm{sharpness}<0.03$. 
Fig. \ref{mocksharp} shows our measured sharpness parameter versus $I_0$ band
magnitude. Fig. \ref{mockhistVI} shows the histograms of the injected and
recovered stars as a function of apparent magnitude in 0.25 mag bins. Our
recovery rate falls to $50\%$ completeness at $I_0 \approx 24.7$ and $V_0
\approx 25.7$ (Fig. \ref{mockhistVI}) and the scatter in the difference between the input and the
recovered magnitudes is a satisfactory $\sigma =\pm~0.1$ mag in the range $24.0 < I_0 <
24.7$, with an offset of 0.03 mag (Fig. \ref{IinVsIout}). We made
  a further set of three tests within Quadrant 1 using 5000 stars within a smaller range of RGB magnitudes: $24<I_0<25$. Colors for the stars of $(V-I)_0$ =
  1.2, 1.4 and 1.7 were tested, with negligible effect on the $50\%$
  completeness limit at $I_0=24.7$. We used the same 4000 artificial stars mentioned above and performed the incompleteness test with Source Extractor version 2.5 \citep{Bertin1996}. The recovery rate fell to $50\%$ completeness at $I_0 \approx 24.7$, in agreement with the test performed with DAOPHOT. 
Spatial incompleteness was not a major issue. We did tests of spatial incompleteness with mock stars added to the images with various density law falloff profiles over the CCD -- and were able to recover the density profiles entirely satisfactorily. The slight photometric offsets between the four chips were small enough that experiments on one quadrant only were deemed
  sufficient.

\section{Foreground and Background Contamination and the CMD}
\label{cmd}

\begin{figure}
\centering
\includegraphics[width=8cm]{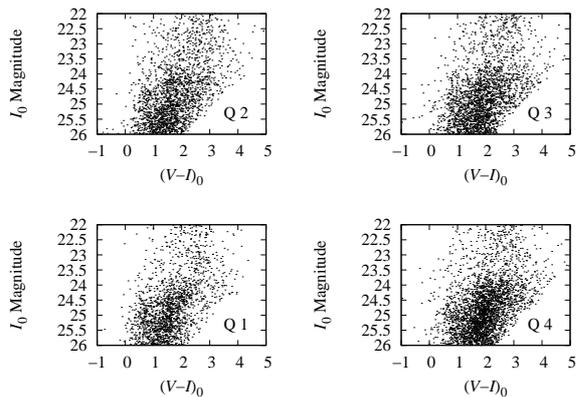}
\caption{NGC 5128 CMDs for each of the four VLT/VIMOS CCD
    quadrants. We count 2365, 2816, 3100, and 3956 point sources in Quadrants
    1, 2, 3, and 4, respectively, within the range of magnitude $21 < I_0 < 27$
    and color $-1 < (V-I)_0 < 5$. Corrections have been made for the aperture
  size, the zeropoint dependencies on color, and the extinction. The TRGB is
  seen at $I_0 \approx 23.9$. Stars above this magnitude are predominantly from
  the Milky Way. Note that Quadrant 4 is the closest of the 4 quadrants to the
  center of NGC 5128, and has the highest source density.}
\label{colormag_quad}
\end{figure}


\begin{figure}
\centering
\includegraphics[width=8cm]{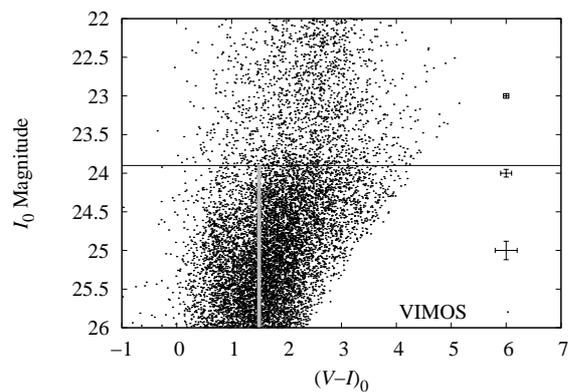}
\caption{NGC 5128 CMD for all four CCD quadrants
    combined. Three rectangular sections
    are shown, two of which are dominated by contaminants to the RGB CMD. The
    cleanest sample of RGB stars in the halo of NGC 5128 are found in the lower
    right section drawn on the CMD. Sources brighter than the TRGB at
    $I_0=23.9$ \citep[we have corrected for extinction]{Harris2010} are field
    stars (upper section) -- we justify this cut in
    Fig. \ref{colormag_fieldstars}. The lower left section (with $(V-I)_0<1.5$)
    is dominated by unresolved background galaxies which thus appear as point
    sources. Justification of the choice of $(V-I)_0=1.5$ for the break between
    background galaxy dominance and RGB dominance of the source counts can be
    seen in Fig. \ref{Q4.VIMOShist24.7}. Error bars for magnitude and color are
    indicated on the righthand side.  }
\label{colormag}
\end{figure}

CMDs of the four quadrants for the final sample of point sources
  found in our VIMOS field are shown in Fig. \ref{colormag_quad}. We find 2365,
  2816, 3100 and 3956 point sources in quadrants 1, 2, 3 and 4, respectively,
  for a total of 12,237 point sources covering a range of magnitude $21 < I_0 <
  27$ and color $-1 < (V-I)_0 < 5$. The TRGB emerges from the field stars at
$I_0 \approx 23.9$, as observed by \citet{Harris2000} and \citet{Harris2010}, and
the $I$ band 50\% completeness limit lies at $I_0 \approx
24.7$. Fig.~\ref{colormag} shows all four quadrants combined into a single CMD.

Our field is located in the distant halo of NGC 5128, and in the next two
subsections contamination of our source counts by foreground stars and
background galaxies is carefully quantified before we isolate a clean sample of
RGB stars from the halo of NGC 5128.

\subsection{Foreground Star Contamination}

\begin{figure}
\centering
\includegraphics[width=8cm]{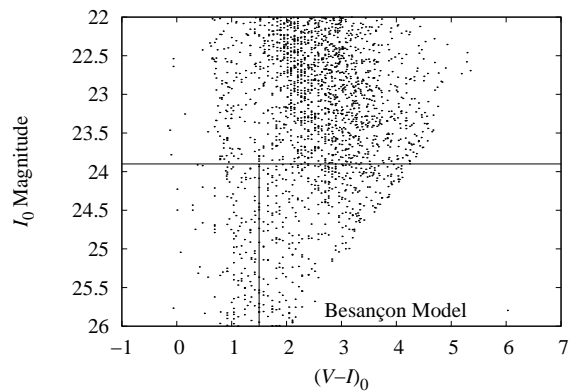}
  \caption{Predictions of the Besan\c{c}on model for field stars
      \citep{Robin2003} in our VIMOS field. Compared to Fig. \ref{colormag},
      the RGB window ($23.9 < I_0 <24.7$ and $1.5 < (V-I)_0 < 3.5$, as
      described in Sec. \ref{contamination}), is mostly uncontaminated by the
      Milky Way's stars as the majority are brighter than NGC 5128's TRGB at
      $I_0=23.9$. Our chosen magnitude limits of $I_0=27.0$ and $V_0=28.2$, based
      on the depth of our VIMOS data, yield the diagonal cut on the righthand
      side of the field star predictions.  }
\label{colormag_fieldstars}
\end{figure}

We estimate the foreground contamination in our fields due to Milky Way stars
via the Besan\c{c}on field star model \citep{Robin2003}\footnote{models
  available at http://model.obs-besancon.fr/} using the Galactic coordinates,
field of view, and depth of our VIMOS data. Fig. \ref{colormag_fieldstars}
shows a CMD from the model, where we use magnitude limits of $I_0=27$ and $V_0
= 28.2$. Most (80\%) of the modeled field stars lie in the magnitude range $21
< I_0 < 23.9$ and are brighter than the TRGB in NGC 5128 (horizontal line in
the figure). Fainter than $I_0 = 23.9$, the number of foreground stars drops
rapidly, as these stars are intrinsically dim, late-type dwarfs seen at
distances of the order of tens of kpc, and the number of such
sources drops rapidly with apparent magnitude.

\begin{figure}
\centering
\includegraphics[width=8cm,angle=-90]{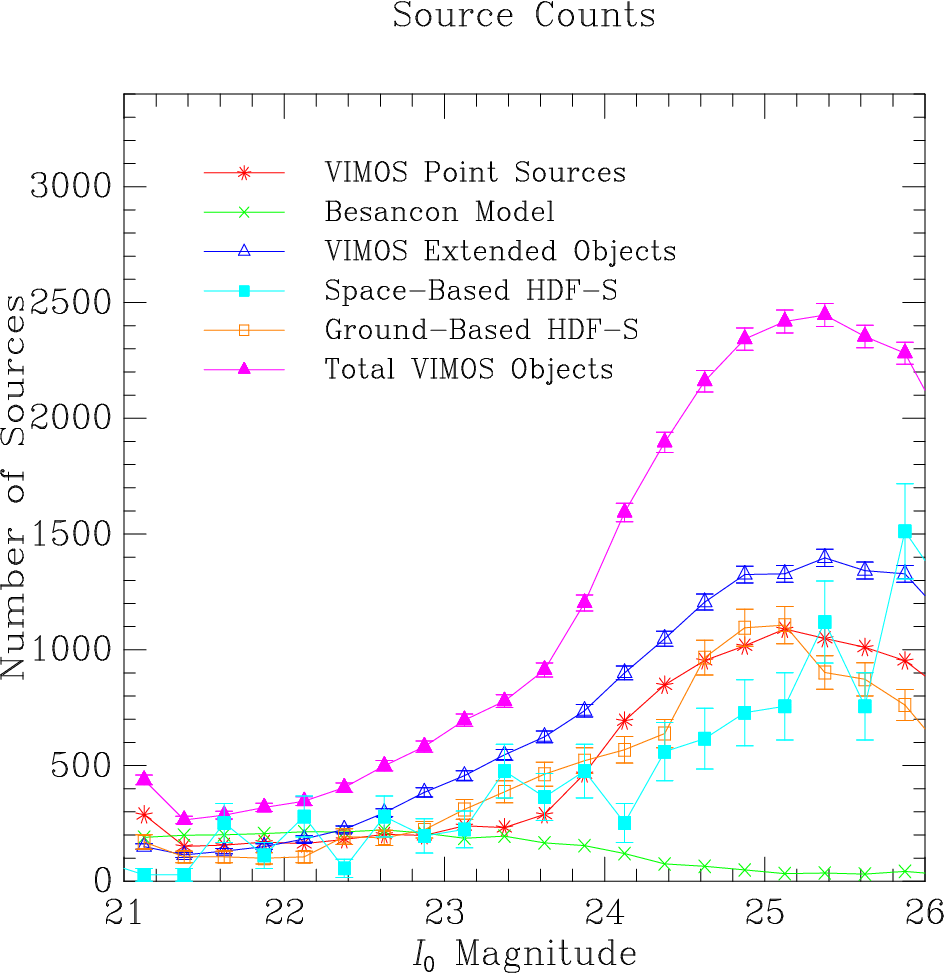}
\caption{ Source counts as a function of apparent $I$ magnitude 
    in our VIMOS data compared to counts in the Besan\c{c}on model of the Milky Way \citep{Robin2003}, space-based Hubble Deep Field South (HDF-S)
    \citep{Williams2000,Casertano2000}, and ground-based HDF-S
    \citep{da_Costa1998}. Both HDF-S source counts have been normalized to the
    field of view of VIMOS. 
Sources in VIMOS have been additionally subdivided into an upper limit of extended
    objects (blue open triangles) and point sources (red asterisks), clearly showing that an excess number of sources in our VIMOS field are detected as compared to the sources of both HDF-S fields. The
    predictions of the Besan\c{c}on model (green crosses) are in good agreement
    with the number of point sources in our VIMOS field down to $I_0 \approx
    23.5$, dimmer than which the point sources in the VIMOS field rises rapidly
    due to a combination of RGB stars in NGC 5128 and rising numbers of
    unresolved, mainly blue, background galaxies. Ground-based HDF-S (orange open squares)
    is a comparative example of faint background galaxy
    observations with similar $50\%$ completeness at $I_0=24.6$ as to our VIMOS study of $I_0=24.7$. The similarity
    between VIMOS point source counts (red asterisks), ground-based extended source counts in the field of NGC 5128 (open blue triangles), and ground-based
    background galaxy counts in the field of HDS-S (open orange squares)
shows that further work is needed to discriminate the two populations
    (cf. Sec. \ref{contamination}). Poisson error bars are added for VIMOS
    total sources and extended objects and for space-based and ground-based HDF-S. Incompleteness has not
    been corrected. Error bars for VIMOS point sources and the Besan\c{c}on model 
    are of the same scale as the plotting symbols and are not included. [See
      the electronic edition of the Journal for a color version of this
      figure.]  }
\label{starcounts}
\end{figure}

Fig. \ref{starcounts} shows the number of foreground stars from the
Besan\c{c}on model (green crosses) as a function of $I_0$ band magnitude, and
is seen to be a good match to the number of point sources in our VIMOS image to
$I_0 \approx 23.5$ (red asterisks). In this apparent magnitude
  range, our completeness is high. Dimmer than this, the number of foreground
  stars decreases to a small fraction of the total number of point sources (red
  asterisks).

\subsection{Background Galaxy Contamination and the RGB Selection Window}
\label{contamination}

For magnitudes dimmer than $I_0 \approx 24$, faint, blue galaxies become an
important contributor to the source counts in our ground-based imaging, as they
will be unresolved. To estimate the contamination rate of such sources in the
CMD in the region of NGC 5128's RGB, we have examined source counts in the space-based
Hubble Deep Field South (HDF-S) \citep{Williams2000,Casertano2000}, as these dim galaxies are well
resolved from space.

We used Version 2 of the final space-based HDF-S catalog \footnote{space-based HDF-S data products available at http://www.stsci.edu/ftp/science/hdfsouth}. The catalog magnitudes were given in the $AB$ system, the field is located high above the Galactic plane, and the detected sources go as deep as $I_0=27.4$. In order to compare with our VIMOS data, we used the offsets in Table 2 of \citet{Frei1994} to convert the $I$ and $V$ band magnitudes from the $AB$ system to the $I$ and $V$ used by VIMOS, namely $I=I(AB)-0.309$ and $V=V(AB)+0.044$. In Fig. \ref{starcounts}, we compare the number of sources in our VIMOS image with the numbers seen in space-based HDF-S as a function of $I_0$ band magnitude, where the space-based HDF-S counts have been scaled by a factor of 28 to match the VIMOS field of view. Dimmer than $I_0=23.5$, the total number of VIMOS sources (lilac filled triangles) rises rapidly compared to space-based HDF-S (cyan filled squares), due to RGB stars in NGC 5128's halo. These can be clearly
seen down to our 50\% completeness limit at $I_0 = 24.7$ (the counts have not
been corrected for completeness in this figure). Using the
  defined sharpness parameter of Sec. \ref{SubSecSharp}, sources in VIMOS have
  also been divided into extended objects (blue open triangles,
  $0.05<\mathrm{sharpness}<1.0$) and point sources (red asterisks). (Note that
dimmer than $I_0 \approx 25$, incompleteness reduces our counts rapidly). HDF-S gives us a means of estimating an upper limit on the number of point sources in
VIMOS which are due to unresolved galaxies rather than to RGB stars in NGC
5128, and thus contaminating our counts. The comparison between our VIMOS field and HDF-S is a preliminary, quantitative comparison that will need more careful analysis in the future.
As we show next, these unresolved galaxies
are predominantly bluer than $(V-I)_0 = 1.5$, and hence mainly blueward of the
NGC 5128 RGB stars, such that a color cut can be used to cleanly remove them.

\begin{figure}
\centerline{\includegraphics[width=13cm]{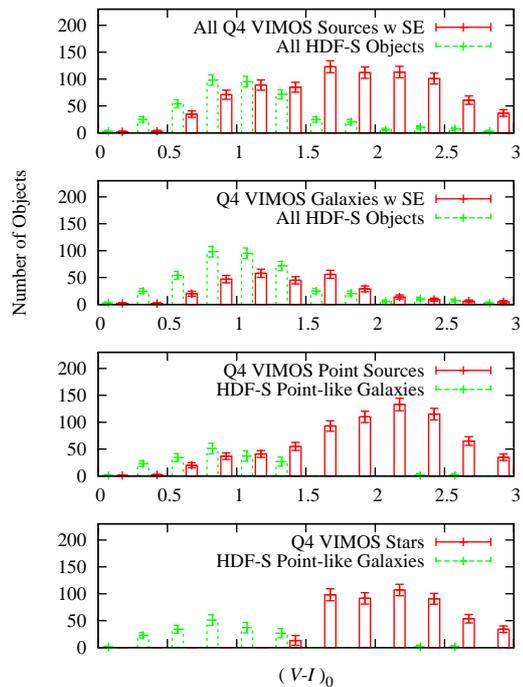}}
\caption{Estimation of the unresolved galaxy contamination rate in the VIMOS
  data with $23.9 < I_0 < 24.7$ as a function of $(V - I)_0$
  color. Poisson distribution errors in number counts are
    shown. Incompleteness has not been corrected for either VIMOS or ground-based HDF-S. 
The histograms are of objects in color bins of width
  $\Delta ((V-I)_0=0.25)$ and are scaled to the field of view for one 36.6
  square arcminute quadrant of our VIMOS data. The first panel from the top shows all sources
  from VIMOS Quadrant 4 as found by Source Extractor version 2.5
  \citep{Bertin1996} (red histogram) and the objects from the ground-based
  observations by \citet{da_Costa1998} of the HST Hubble Deep Field South
  (HDF-S) \citep{Williams2000} (dashed green histogram). 
The second panel from the top shows extended sources in
  VIMOS Quadrant 4 found using Source Extractor (red histogram) and again the
  objects from the ground-based observations of HDF-S (dashed green
  histogram). 
The third panel from the top shows the point sources from our VIMOS data as the red histogram. 
The dashed green histogram shows the difference between
  the histograms in the second panel.
The last panel shows two difference histograms. The red
  histogram shows the difference between the two histograms in the first panel.
The dashed green histogram is the same as that in the third panel. 
See Sec. \ref{contamination} for a full discussion of this figure. 
[See the electronic
    edition of the Journal for a color version of this figure.]
}
\label{Q4.VIMOShist24.7}
\end{figure}

We show this by comparing our VIMOS data with a second field,
  the ground-based observations by \citet{da_Costa1998} of HDF-S \citep{Williams2000}, which are represented in
  Fig. \ref{starcounts} by the open orange squares (scaled to match the total
  four VIMOS quadrants' field of view, although not corrected for
  incompleteness). The camera used was SuSI2 \citep{DOdorico1998}, mounted on
  the f/11 Nasmyth focus A of the ESO 3.58~m New Technology Telescope. The field
  covered 24.8 square arcminute and integrated magnitudes (corrected for
  reddening using \citet{Schlegel1998} $E(B-V) = 0.027$, yielding the
  extinction correction $A_I=0.04$ mag) were obtained using Source Extractor
  version 2.5 \citep{Bertin1996}. \citet{da_Costa1998} determined the data were $50\%$ complete to $I_0 =
  24.6$ by comparing the region of overlap between the ground-based and space-based HDF-S
  catalogs, and had a stellar classification limit of $I_0 \la 22.0$ (based on
  comparing counts from the stellar classification of Source Extractor and Milky Way model
  predictions). We did not make corrections for incompleteness since both VIMOS
  and ground-based HDF-S data share similar $50\%$ completeness limits at
  $I_0=24.7$ and 24.6, respectively. In the following we used the same
software on our VIMOS data to facilitate as direct a comparison as possible
with HDF-S, in particular to be able to compare the integrated magnitudes of
the extended objects. In Fig. \ref{Q4.VIMOShist24.7}, we plot histograms by
$(V-I)_0$ color of sources in HDF-S and VIMOS Quadrant 4, in the magnitude
range $23.9 < I_0 < 24.7$ in bins of width $\Delta((V-I)_0) = 0.25$. Note that
we normalize HDF-S counts to the area of VIMOS Quadrant 4 (i.e. 36.6 square
arcminute).

The first panel from the top of Fig. \ref{Q4.VIMOShist24.7} shows the
  histograms by $(V-I)_0$ color of the extended and point source objects found
  in VIMOS Quadrant 4 (red histogram), compared to the numbers found in ground-based HDF-S
  (dashed green histogram). The background galaxies in HDF-S have a color
  distribution which peaks at $(V-I)_0 \approx 1.0$, whereas the peak of the
  VIMOS data lies at $(V-I)_0 \approx 1.8$ due to the RGB stars in the
  field. The red histogram shows a clear
  excess of red sources ($(V - I)_0 > 1.5$) due to RGB stars in NGC 5128 in the
  VIMOS data compared to HDF-S. In the second panel from the top, we show sources classified as resolved galaxies (extended sources)
  in VIMOS Quadrant 4 (red histogram), compared to the galaxies found in HDF-S
  (dashed green histogram). The two histograms are a fair match which indicates
  we are seeing the same (resolved) galaxy population in both fields. The red histogram in the third panel from the top shows all point sources from our VIMOS data. These sources consist of NGC 5128 RGB stars, unresolved background galaxies, and foreground stars. The dashed green histogram in the third panel 
  contains the sources which remain after subtracting the two histograms from the second panel, namely the red VIMOS histogram and
  the dashed green HDF-S histogram of the second panel. This dashed green subtracted histogram of the third panel serves as an indicator of the color distribution
  of galaxies in HDF-S, which 
we interpret as the
  residual, unresolved galaxies in our VIMOS data. The two histograms of the third panel show how many of our VIMOS point sources could possibly be contaminants due to unresolved background galaxies. VIMOS point sources with $V-I<1.5$ are dominated by unresolved background galaxies. 
In the last panel, we show two difference histograms. The red histogram gives an indication of the color distribution of sources in VIMOS Quadrant
  4 interpreted as RGB stars in NGC 5128 (plus a minor contaminant of Milky Way
  foreground stars) 
and is the difference
  histogram 
produced by subtracting the red and the dashed green histograms of
  the first panel. 
The dashed green difference histogram in the last panel is the same as in the third panel. The last panel shows that redder
  than $(V-I)_0 = 1.5$, contamination by unresolved galaxies in the RGB is very
  small (of order a few percent).

We conclude from this comparison with source counts in HDF-S, that we can
reliably separate out RGB stars in NGC 5128's halo redward of $(V-I)_0 = 1.5$:
but that blueward of this limit the counts become quickly swamped by unresolved
galaxies. Hence, our final window for RGB selection is
$23.9<I_0<24.7$ and $1.5<(V-I)_0<3.5$. We should point out that we
have not made corrections for the incompleteness levels in either catalog since VIMOS and ground-based HDF-S share similar $50\%$ completeness levels at $I_0=24.7$ and 24.6, respectively. 
Completeness affects both the faint and the red end of the considered magnitude range (overestimating faint blue point sources due to contamination by blue, background galaxies and underestimating red sources due to the $V$ band limit), and a more quantitative investigation of this issue will be presented in a future contribution. This is a
conservative approach, as VIMOS is likely to be more affected by incompleteness
than HDF-S, boosting the number of sources in the VIMOS field relative to
HDF-S, and making the color cut at $(V-I)_0 = 1.5$ more effective still in
reducing the contamination of the RGB by unresolved background galaxies. The possible relaxation of such a color cut will be part of our future study.

In summary, to isolate RGB stars in NGC 5128's outer halo, we choose the window
$23.9 < I_0 < 24.7$ and $1.5<(V - I)_0<3.5$ in the CMD. These selection limits
are driven by the apparent magnitude of the TRGB at $I_0=23.9$, the
completeness limit of $50\%$ recovered sources in both bands, at $I_0 = 24.7$,
and avoidance of predominantly blue, unresolved background galaxies $(V-I)_0 >
1.5$. These criteria yield our final sample of 1581 point sources in our VIMOS
data which we interpret as RGB stars in NGC 5128's outer halo.


\section{The Metallicity and Density-Falloff of NGC 5128's Stellar Halo}
\label{mdf}

\begin{table}
\caption{Raw numbers of point sources observed with HST and VLT/VIMOS, at the given projected distance (in kpc) from the center of NGC 5128,
    within the defined CMD overlap region $1.5 < (V-I)_0 < 2.3$ and $-3.0 >
    M_{\mathrm{bol}}>-3.22$. Due to low counts in the metal-poor VIMOS data,
    all sources have been binned together and included in the average distance
    bin of 65 kpc.}
\label{overlap}
\begin{center}
\begin{tabular}{|ccc|}
\hline\hline
Distance (kpc) & [M/H]~$>-0.7$& [M/H]~$<-0.7$\\
\hline
9 & 120 & 8 \\
24 & 185 & 15 \\
35 & 66 & 10 \\
52 & 64 & ...\\
56 & 69 & ...\\
61 & 88 & ...\\
65 & 62 & 39\\
70 & 27 & ...\\
\hline
\end{tabular}
\end{center}
\end{table}

\subsection{MDFs}
\begin{figure}
\centering
\includegraphics[width=8cm]{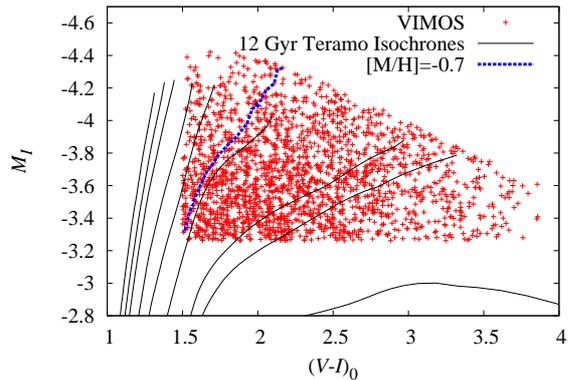}
\caption{Stellar evolutionary isochrones superimposed over the CMD for our
  65~kpc field against absolute magnitude $M_I$ and $(V-I)_0$ color. The models
  assume RGB stars of age 12 Gyr. Solid black lines are the
  $\alpha$-enhanced isochrones from
  \citet{Pietrinferni2004,Pietrinferni2006}; from left to right, the
  metallicities are [M/H] $= -2.27, -1.79, -1.49, -1.27, -0.96, -0.66, -0.35,
  -0.25, 0.06$. The dashed blue line marks [M/H]$ = -0.7$, the cut at which we
  divide between metal-poor and metal-rich objects. The TRGB
    begins at $M_I=-4.05$ \citep{Rizzi2007}.  }
\label{cmd_isochrones}
\end{figure}

Metallicities for our sample of RGB stars were estimated using
  the procedure described in \citet{Harris2000} with the additional
  improvements described in \citet{Harris2002}, by interpolation of $(V-I)_0$
  colors in the BaSTI (also referred to as Teramo) 12~Gyr
  isochrones\footnote{see http://albione.oa-teramo.inaf.it/ to retrieve the
    evolutionary track isochrones}
  \citep{Pietrinferni2004,Pietrinferni2006}. We chose the $\alpha$-enhanced
  BaSTI isochrones because the enhancement fits the old stellar halo population
  better than scaled-solar element ratios, as shown by \citet{Rejkuba2011} for
  their 40 kpc NGC 5128 field. The average
  $\alpha$-enhancement was $[\alpha/\mathrm{Fe}]\sim0.4$, although
  for more details, see \citet{Pietrinferni2006} and references within.
Fig. \ref{cmd_isochrones} shows the 12 Gyr isochrones overplotted on our NGC
5128 CMD in $(V-I)_0$ versus absolute magnitude $M_I$, where we adopted the
distance modulus of $(m-M)_0(\mathrm{TRGB})=27.91\pm0.08$, calculated by
averaging together the distance to NGC 5128 by several TRGB studies
\citep{Harris2010}. Bolometric magnitudes $M_{\mathrm{bol}}$,
  were estimated for the stars following \citet{Harris2000},
\begin{displaymath}M_{\mathrm{bol}}\equiv(M_V-\mathrm{BC}) = M_I + (V-I)_0 - \mathrm{BC}~~~,\end{displaymath}
where BC is the bolometric correction estimated as function of color through a
rough linear two-segment curve,
   \begin{displaymath}
\mathrm{BC}=
\left\{
\begin{array}{c l}     
1.0583 \times (V-I)_0 -0.4348,&~~(V-I)_0\le2.3 \\
1.4441 \times (V-I)_0 -1.3210,&~~(V-I)_0>2.3~~~.
\end{array}\right.
   \end{displaymath}
An internal uncertainty of $\pm 0.2$ in the deduced [M/H] was
estimated from the combined $V_0$ and $I_0$ band photometric errors. For a
direct comparison with the previously studied NGC 3379 halo field in which a
transition in metallicity was found \citep{Harris2007}, we adopted their
metallicity cut at [M/H]$ = -0.7$ as the division between metal-poor and
metal-rich sources (dotted blue curve in Fig. \ref{cmd_isochrones}).

\begin{figure}
\centering
\includegraphics[width=8cm]{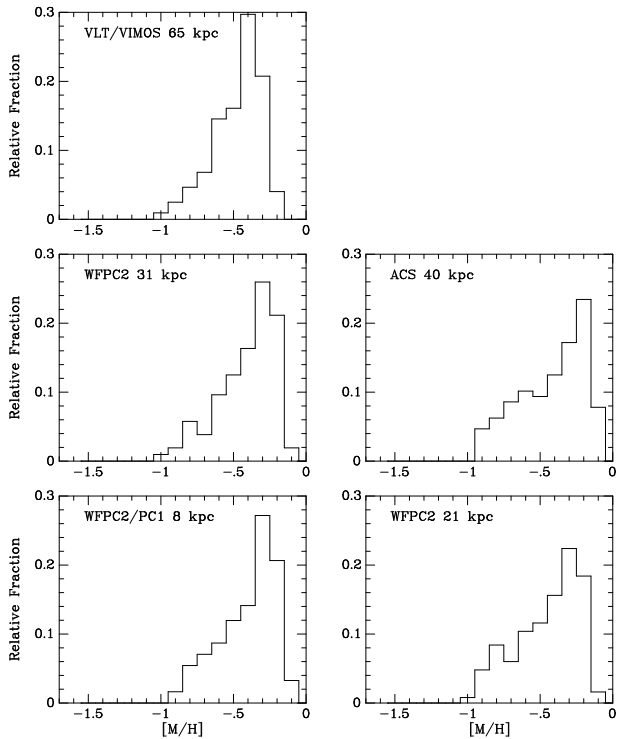}
\caption{Normalized MDFs in [M/H] for five stellar halo fields
    in NGC 5128, our 65 kpc VLT/VIMOS field and four HST fields: 31 kpc
    \citep{Harris2000}, 40 kpc \citep{Rejkuba2005}, 8 kpc \citep{Harris2002},
    and 21 kpc \citep{Harris1999}. The data included are in the overlapping
    region of $1.5 < (V-I)_0 < 2.3$ and $-3.0 > M_{\mathrm{bol}}>-3.22$. Neither VIMOS nor HST data has been corrected for incompleteness. 
}
\label{fehdist}
\end{figure}

We have also computed metallicities and bolometric magnitudes for stars in halo
fields of NGC 5128 imaged with HST closer to the galaxy center than VIMOS, at
distances of 8 kpc \citep{Harris2002}, 21 kpc \citep{Harris1999}, 31 kpc
\citep{Harris2000}, and 40 kpc \citep{Rejkuba2005}. As these are observations
from space, the background galaxies can easily be found and removed, but these
galaxies are a minor component in the source counts in any case, because the
surface density of RGB stars is so high compared to background galaxies, quite
unlike our VIMOS field.

We have carefully determined over what range of color and bolometric magnitude
our VIMOS study and the HST studies completely overlap in sampling of the RGB
region. This turns out to be in the color range $1.5 < (V-I)_0 <
  2.3$ and bolometric magnitude range $-3.0 >
  M_{\mathrm{bol}}>-3.22$. Metallicities of the RGB stars meeting these
  criteria in the HST fields have been computed using the same method as we
  used for our 65 kpc field (namely the Teramo isochrones and the cut between
  metal-rich and metal-poor stars at [M/H] $ = -0.7$). Metal-rich stars are preferentially missed because of increasing incompleteness as a function of color, although we do not correct for incompleteness in this comparison as we are especially interested in the metal-poor region. We compare the raw
  number counts of the metal-rich and metal-poor data within the overlap region
  in Table \ref{overlap}. While this procedure reduced our basic sample
considerably, it has allowed us to examine the density falloff of both
metal-rich and metal-poor stars over almost an order of magnitude in distance
range, from $\approx 8$ to $\approx 65$ kpc.

The normalized MDF of our 65 kpc field (VIMOS) is compared with the deep HST
fields of \citet{Harris2002}, \citet{Harris1999}, \citet{Harris2000}, and
\citet{Rejkuba2005} (8, 21, 31, and 40 kpc, respectively) in Fig. \ref{fehdist}
      over the common region of $1.5 < (V-I)_0 < 2.3$ and $-3.0 >
        M_{\mathrm{bol}}>-3.22$. The MDFs of HST and our 65 kpc field are
      quite similar, indicating that our outer halo field is not obviously more
      metal-poor than the halo fields closer towards the center of NGC 5128.
      To large measure, the absence of RGB stars more metal-poor than [M/H]
      $\simeq -1$ is due to our color selection criteria (see
      Fig.~\ref{cmd_isochrones}), which artificially cut off almost all stars
      to the blue side of the [M/H] $= -0.96$ isochrone line.  In short, we
      find that the predominantly metal-rich nature of the NGC 5128 halo
      continues on outward to at least 65 kpc or $\sim 12 R_\mathrm{eff}$.

Using data of similar quality to ours, with seeing of
$0.57\arcsec - 0.76\arcsec$, and reaching to similar depth,
\citet{Crnojevic2013} found clear metal-rich and metal-poor components to
  the outer halo of NGC 5128. 
They determined metallicities for their RGB stars, finding
median abundances in the range [Fe/H]$_\mathrm{med} = -0.9$ to $-1.0$,
which vary little over the fields observed, despite the large range of radii
covered. Taking [M/H]$=$[Fe/H]$+$[$\alpha/$H] into account, their
  MDFs are in reasonable agreement with ours around the median abundance, given
  the uncertainties and different methods, but they found long tails in the MDF
  to low metallicity. Our color cut of $(V-I)_0 > 1.5$ means that low
metallicity stars ([M/H] $\la -1$) have been explicitly removed from our sample
(as they overlap in the CMD with blue, unresolved galaxies, see Fig.
\ref{cmd_isochrones}). We did not investigate the metallicity of
  sources in this region because we consider it to be dominated by
  contaminating galaxies, bluer than $(V-I)_0 = 1.5$. 
  Comparison with the MDFs of this paper and those of \citet{Crnojevic2013} is
  complicated by this color cut: nevertheless, within the overlap between our
  RGB selection criteria (i.e. redward of $(V-I)_0 = 1.5)$, we find a fairly
  similar peak metallicity and negligible gradient across the field, in good
  agreement with the main findings of \citet{Crnojevic2013}.

Space-based data of these
  outer regions confirmed with \citet{Crnojevic2013} that the MDF does reach bluer than $(V-I)_0 = 1.5$, with blue tails more metal-poor than [M/H]$=-2.0$ \citep{Rejkuba2014}. The distant NGC 5128 halo regions studied by \citep{Rejkuba2014} revealed that a gradient exists in the MDF past our 65 kpc field, although variations in metallicity exist among their fields. The average metallicity out to 140 kpc remained above [M/H]$=-1$.

Archival HST data for fields closer in to the center of the galaxy are
available, and we intend to follow up the background contamination rates by
estimating the source density of galaxies directly behind the galaxy itself, as
an alternative to our comparison with the ground-based HDF-S study, as used in
Sec. \ref{cmd}. We intend also to look at the distribution of the
  substantial number of stars with $V-I<1.5$ found by \citet{Crnojevic2013}
  within this future project.


\subsection{Surface Brightness Profiles}
\label{sbprofile}
\begin{figure}
\centering
\includegraphics[width=8cm,angle=-90]{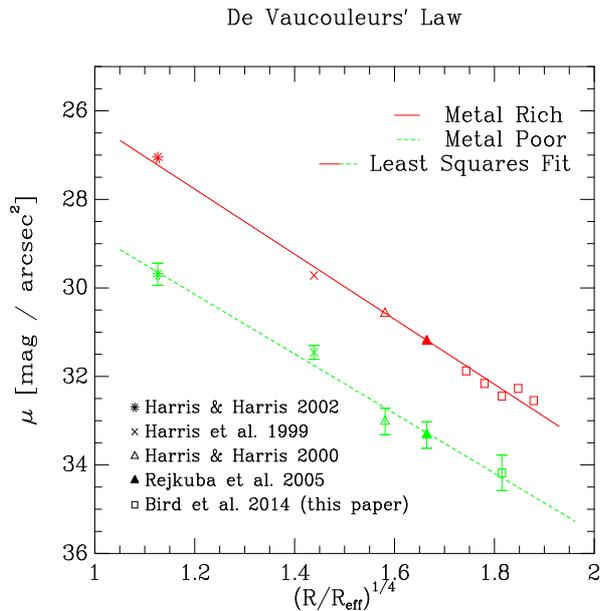}
\caption{De Vaucouleurs' law fits (red and dashed green
    lines) to the surface brightness $\mu$ profile of metal-rich (red symbols
    with red fit line) and metal-poor (green symbols with dashed green fit
    line) stars in the halo of NGC 5128. We show $\mu$ in magnitudes per square
    arcsecond of the NGC 5128 RGB stars as a function of
    $(R/R_\mathrm{eff})^{1/4}$, where $R$ is the ellipticity corrected distance
    from the center of the galaxy and $R_\mathrm{eff}=5.6$ kpc
    \citep{Dufour1979}. Within this plane, $\mu$ versus
    $(R/R_\mathrm{eff})^{1/4}$, we use the least squares method to fit the data
    and solve for the slope $a$ and y-intercept $\mu_0$ of de Vaucouleurs'
    profile \citep{de_Vaucouleurs1948} $\mu = a~(R/R_\mathrm{eff})^{1/4} + \mu_0$,
    where $R_\mathrm{eff}$ is fixed at 5.6 kpc. We correct for incompleteness
    by doubling the number of sources in our VIMOS field. 
Note that we have had to use only
    one distance bin for the metal-poor stars in VIMOS (due to small number
    statistics) whereas the metal-rich are numerous enough to be placed into 5
    distance bins across the VIMOS field. Poisson distribution errors are
    included, although none appear for the red metal-rich symbols (fit by the
    red line) because the errors are of order of the same size as the symbol
    size. 
See Sec. \ref{sbprofile} for a full discussion of this figure. [See the electronic edition of the Journal for a color version of
      this figure.]  }
\label{vaucouleurs}
\end{figure}

In Fig. \ref{vaucouleurs}, we show the density falloff of
  metal-rich ([M/H]$ > -0.7$) and metal-poor stars ([M/H]$ < -0.7$) in the halo
  of NGC 5128, using the same overlap sample as plotted in Fig. \ref{fehdist},
  for which raw numbers are given in Table \ref{overlap}. To calculate the
  effective sampling area of our VIMOS images, we evenly spread mock sources
  over each quadrant and used the same DAOPHOT procedure as described in
  Sec. \ref{SubSecSharp} to detect the sources. We used these detections to
  compute the effective sampling area on the quadrants of five radially
  concentric bins centered on NGC 5128.
In Fig. \ref{vaucouleurs}, we have multiplied the VIMOS counts by a factor of
two in order to account for $50\%$ incompleteness. We also show least squares
de Vaucouleurs' law fits to the falloff luminosity $\mu \propto
(R/R_\mathrm{eff})^{1/4}$ \citep{de_Vaucouleurs1948}, where $\mu$ is the
surface luminosity, $R$ is the ellipticity corrected distance from the center
of the galaxy (with an ellipticity correction of $b/a=0.77$, this gives
corrected radii for the HST fields of 9, 24, 35, and 43 kpc), and
$R_\mathrm{eff}=5.6$~kpc \citep{Dufour1979}. We have converted star counts in
the fields into surface brightnesses by assuming a typical apparent magnitude
of $ I_0 = 24.5$ for the RGB sample since they lie in the overlap range of
$24.6>I_0>24.35$ ($-3.0 > M_{\mathrm{bol}}>-3.22$). We fix the value of
$R_\mathrm{eff}$ to 5.6 kpc and perform a least squares fit to the data in the
$\mu$ versus $(R/R_\mathrm{eff})^{1/4}$ plane. We find that the same de Vaucouleurs'
law, $\mu = a~(R/R_\mathrm{eff})^{1/4} + \mu_0$, where $a$ is the slope and $\mu_0$ is
the y-intercept, is a good fit to both the metal-rich and metal-poor stars,
with slopes of $a=7.3\pm0.2$ and $6.7\pm0.4$, respectively.

We find definite metal-rich and metal-poor halo populations, although we see no evidence of a transition from the dominating metal-rich halo to
metal-poor halo in the radial falloff in density of the RGB stars in our VIMOS
field and the HST fields. If the transition exists in NGC 5128, we expected
the slope of the metal-rich stellar falloff to become much steeper than
$R^{1/4}$ near $10~R_\mathrm{eff} \approx
    60$ kpc or $(R/R_\mathrm{eff})^{1/4}=1.8$ in Fig. \ref{vaucouleurs}, 
revealing an underlying
    metal-poor component, as seen in the galaxies M31 \citep{Kalirai2006} and
    NGC 3379 \citep{Harris2007}. On the contrary, the populations of both
    metal-rich and metal-poor stars are found out to
    $(R/R_\mathrm{eff})^{1/4}=1.9$ or $13~R_\mathrm{eff} \approx 70$ kpc and
    exhibit constant falling $\mu$ profiles. 
 Any such transition must lie still further out.  However, the
absence of such a transition (so far) might go along with the history
of NGC 5128 as a merger remnant.  Age dating of the outer-halo stars
in the 40 kpc field \citep{Rejkuba2011} and of the globular clusters \citep{Woodley2010ApJ} consistently indicates that perhaps $1/4$ of the
stars were formed in some kind of merger event $3 - 4$ Gyr ago, with the
rest of the halo being classically ``old'' (~$>$ 10 Gyr). 
A merger on this scale could dynamically shuffle the orbits well out into
the halo, so even if it originally had a transition region at $\sim 10~R_\mathrm{eff}$ as in M31 and NGC 3379, it may have been diluted or
moved further outward.

Despite a significant difference between the
  current study and that of \citet{Crnojevic2013} in the selection criteria for
  RGB stars, our density falloffs for halo stars are in agreement. Our selection criteria are
$23.9 < I_0 < 24.7$ and $1.5<(V-I)_0<3.5$, set by the magnitude of the TRGB
($I_0=23.9$), our 50\% completeness limit ($I_0 = 24.7$), and the color regime
in which we consider RGB stars to dominate unresolved galaxies in the source
counts. \citet{Crnojevic2013} used a 7$-$sided polygon in the CMD to select RGB
stars (their Fig. 4), a procedure which differs substantially from our RGB
selection in going deeper to (extinction corrected) $I_0 = 25.0,$ and extending
to the blue as far as (dereddened) $(V-I)_0 = 1.0$ (at the faint limit) and to
$(V-I)_0 = 1.4$ at the TRGB magnitude.

\citet{Crnojevic2013} are well aware of potential contamination in their RGB sample,
and show two profiles for the density falloff of the stellar halo in NGC 5128,
based on two methods to correct for background galaxy contamination. The most
conservative option is to make use of source counts in their outermost field,
assuming that these are all background galaxies (at most), and subtracting
these counts from fields further in. Doing this, \citet{Crnojevic2013} find
that the density falloff of the NGC 5128 halo can be quite well fit along the
major axis by a de Vaucouleurs' profile to 75 kpc (see their Fig. 10). We have
explicitly fit a de Vaucouleurs' profile to the \citet{Crnojevic2013} stellar
density falloff along the major axis (solid symbols in their Fig. 10), finding
a slope of $7.1 \pm 1.5$, consistent with our own measure of $7.3 \pm 0.2$ from
our combined metal-rich VIMOS and HST data and covering a much wider range of
galactocentric radii. We conclude that our conservative option of cutting away
sources with $(V - I)_0 < 1.5$ to avoid background galaxy contamination, and
the conservative option of \citet{Crnojevic2013} of estimating background
contamination from their outermost field, give a consistent picture for the
density falloff of NGC 5128's outer halo. Taking into consideration their more distant fields, \citet{Rejkuba2014} confirm that a de Vaucouleurs' profile fits the density falloff best out to $\approx56$ kpc, although their more distant fields along the major axis are best fit by a power law out to 140 kpc. 

\section{Summary and Conclusions}
\label{Conclusions}

We have used VIMOS on VLT to image for almost 6 hours in a 185 kpc$^2$ halo
field located 65 kpc ($\approx12~R_\mathrm{eff}$) from the
center of \object{NGC 5128} / \object{Centaurus A}. We obtained
  photometry in $I$ and $V$ of NGC 5128 RGB stars reliably to 0.8 mag below the
  TRGB at $I_0 = 23.9$ \citep{Harris2010}.

We found that the density of RGB stars $60-70$~kpc from the center of NGC 5128
is sufficiently low that contamination by unresolved background galaxies is a
serious issue. By comparing with the source density of galaxies in the
\citet{da_Costa1998} ground-based study of Hubble Deep Field South
\citep{Williams2000}, we showed that in the magnitude and color
  window $23.9 < I_0 <24.7$ and $1.5 < (V-I)_0 < 3.5$, a clean sample ($>99\%$
  pure) of RGB stars can be isolated. We used stellar isochrones to measure the
  metallicity of the stars and found a broad spread of metallicity including
  both a metal-rich ([M/H]$>-0.7$) and a metal-poor ([M/H]$<-0.7$) halo
  component, similar to what has been seen in other NGC 5128 halo fields
  \citep{Harris2002,Harris1999,Harris2000,Rejkuba2005}, which have been
  observed with HST closer to the galaxy center.

We have searched for a ``transition'' in the metallicity distribution around
$10-12~R_\mathrm{eff}$ from the center, in which the metal-rich component of
the halo would drop away rapidly in number density relative to the metal-poor
halo. This transition has been found in two morphologically different
galaxies, the spiral galaxy \object{M31} \citep{Kalirai2006} and giant
elliptical \object{NGC 3379} \citep{Harris2007}. We found no such transition in 
NGC 5128 in agreement with previous literature results based on a similar dataset, but rather see that both metal-rich and metal-poor halo stars follow the same de Vaucouleurs' profile over radii from 8 kpc to 70 
kpc. Our derived MDF and stellar density falloff are in agreement with the similar outermost halo fields of NGC 5128 studied by \citet{Crnojevic2013} and \citet{Rejkuba2014}.

If such a transition takes place in the halo of NGC 5128,
beyond~$\approx~75$~kpc, the contamination issues of background galaxies will
continue and will make it difficult to find metal-poor stars using ground-based
observations of NGC 5128's outer halo --- space-based data will likely be
required to work yet further out for the background galaxy separation and to
probe the low metallicity tail of the MDF. NGC 5128 presents a unique field to investigate the limits of ground-based faint star detections and background galaxy contamination. We plan to exploit the extensive space- and ground-based observations of NGC 5128 to fully investigate the effects of background galaxy contamination on deep field studies of stellar populations. Such a study is especially relevant for further deep ground-based studies with the 8.2 m VLT, 10 m-class telescopes, and future telescopes such as the Thirty Meter Telescope and the 40 m European Extremely Large Telescope.

\begin{acknowledgements}
SB thanks the Wihuri Foundation for continued funding throughout the project,
the Finnish Cultural Foundation for funding the last part of the research, and
the University of Turku Foundation and the StarryStory Foundation for funding
travel related to the project. We are grateful for the help provided by Marina
Rejkuba during the telescope application period and analysis of the 40~kpc NGC
5128 halo field, to Glen Mackie for very useful discussions, and for the care and interest taken by the referee during the reviewal process. This work has
made use of BaSTI web tools.
\end{acknowledgements}

\bibliographystyle{apj} 
\bibliography{/home/sarah/bibfile.bib} 

\begin{thebibliography}{}
\expandafter\ifx\csname natexlab\endcsname\relax\def\natexlab#1{#1}\fi

\bibitem[{{Beasley} {et~al.}(2008){Beasley}, {Bridges}, {Peng}, {Harris},
  {Harris}, {Forbes}, \& {Mackie}}]{Beasley2008}
{Beasley}, M.~A., {Bridges}, T., {Peng}, E., {et~al.} 2008, \mnras, 386, 1443

\bibitem[{{Bertin} \& {Arnouts}(1996)}]{Bertin1996}
{Bertin}, E., \& {Arnouts}, S. 1996, \aaps, 117, 393

\bibitem[{{Bullock} \& {Johnston}(2005)}]{Bullock2005}
{Bullock}, J.~S., \& {Johnston}, K.~V. 2005, \apj, 635, 931

\bibitem[{{Cardelli} {et~al.}(1989){Cardelli}, {Clayton}, \&
  {Mathis}}]{Cardelli1989}
{Cardelli}, J.~A., {Clayton}, G.~C., \& {Mathis}, J.~S. 1989, in IAU Symposium,
  Vol. 135, Interstellar Dust, ed. L.~J. {Allamandola} \& A.~G.~G.~M.
  {Tielens}, 5P

\bibitem[{{Casertano} {et~al.}(2000){Casertano}, {de Mello}, {Dickinson},
  {Ferguson}, {Fruchter}, {Gonzalez-Lopezlira}, {Heyer}, {Hook}, {Levay},
  {Lucas}, {Mack}, {Makidon}, {Mutchler}, {Smith}, {Stiavelli}, {Wiggs}, \&
  {Williams}}]{Casertano2000}
{Casertano}, S., {de Mello}, D., {Dickinson}, M., {et~al.} 2000, \aj, 120, 2747

\bibitem[{{Crnojevi{\'c}} {et~al.}(2013){Crnojevi{\'c}}, {Ferguson}, {Irwin},
  {Bernard}, {Arimoto}, {Jablonka}, \& {Kobayashi}}]{Crnojevic2013}
{Crnojevi{\'c}}, D., {Ferguson}, A.~M.~N., {Irwin}, M.~J., {et~al.} 2013,
  \mnras, 432, 832

\bibitem[{{da Costa} {et~al.}(1998){da Costa}, {Nonino}, {Rengelink}, {Zaggia},
  {Benoist}, {Erben}, {Wicenec}, {Scodeggio}, {Olsen}, {Guarnieri}, {Deul},
  {D'Odorico}, {Hook}, {Moorwood}, \& {Slijkhuis}}]{da_Costa1998}
{da Costa}, L., {Nonino}, M., {Rengelink}, R., {et~al.} 1998, ArXiv
  Astrophysics e-prints, arXiv:astro-ph/9812105

\bibitem[{{de Vaucouleurs}(1948)}]{de_Vaucouleurs1948}
{de Vaucouleurs}, G. 1948, Annales d'Astrophysique, 11, 247

\bibitem[{{de Vaucouleurs} {et~al.}(1991){de Vaucouleurs}, {de Vaucouleurs},
  {Corwin}, {Buta}, {Paturel}, \& {Fouqu{\'e}}}]{de_Vaucouleurs1991}
{de Vaucouleurs}, G., {de Vaucouleurs}, A., {Corwin}, Jr., H.~G., {et~al.}
  1991, {Third Reference Catalogue of Bright Galaxies. Volume I: Explanations
  and references. Volume II: Data for galaxies between 0$^{h}$ and 12$^{h}$.
  Volume III: Data for galaxies between 12$^{h}$ and 24$^{h}$.} (Springer)

\bibitem[{{D'Odorico} {et~al.}(1998){D'Odorico}, {Beletic}, {Amico}, {Hook},
  {Marconi}, \& {Pedichini}}]{DOdorico1998}
{D'Odorico}, S., {Beletic}, J.~W., {Amico}, P., {et~al.} 1998, in Society of
  Photo-Optical Instrumentation Engineers (SPIE) Conference Series, Vol. 3355,
  Optical Astronomical Instrumentation, ed. S.~{D'Odorico}, 507--511

\bibitem[{{Dufour} {et~al.}(1979){Dufour}, {Harvel}, {Martins}, {Schiffer},
  {Talent}, {Wells}, {van den Bergh}, \& {Talbot}}]{Dufour1979}
{Dufour}, R.~J., {Harvel}, C.~A., {Martins}, D.~M., {et~al.} 1979, AJ, 84, 284

\bibitem[{{Frei} \& {Gunn}(1994)}]{Frei1994}
{Frei}, Z., \& {Gunn}, J.~E. 1994, \aj, 108, 1476

\bibitem[{{Graham}(1979)}]{Graham1979}
{Graham}, J.~A. 1979, \apj, 232, 60

\bibitem[{{Gregg} {et~al.}(2004){Gregg}, {Ferguson}, {Minniti}, {Tanvir}, \&
  {Catchpole}}]{Gregg2004}
{Gregg}, M.~D., {Ferguson}, H.~C., {Minniti}, D., {Tanvir}, N., \& {Catchpole},
  R. 2004, \aj, 127, 1441

\bibitem[{{Harris} \& {Harris}(2000)}]{Harris2000}
{Harris}, G.~L.~H., \& {Harris}, W.~E. 2000, \aj, 120, 2423

\bibitem[{{Harris} {et~al.}(1999){Harris}, {Harris}, \& {Poole}}]{Harris1999}
{Harris}, G.~L.~H., {Harris}, W.~E., \& {Poole}, G.~B. 1999, \aj, 117, 855

\bibitem[{{Harris} {et~al.}(2010){Harris}, {Rejkuba}, \& {Harris}}]{Harris2010}
{Harris}, G.~L.~H., {Rejkuba}, M., \& {Harris}, W.~E. 2010, \pasa, 27, 457

\bibitem[{{Harris} \& {Harris}(2002)}]{Harris2002}
{Harris}, W.~E., \& {Harris}, G.~L.~H. 2002, \aj, 123, 3108

\bibitem[{{Harris} {et~al.}(2007{\natexlab{a}}){Harris}, {Harris}, {Layden}, \&
  {Stetson}}]{Harris2007-ngc3377}
{Harris}, W.~E., {Harris}, G.~L.~H., {Layden}, A.~C., \& {Stetson}, P.~B.
  2007{\natexlab{a}}, \aj, 134, 43

\bibitem[{{Harris} {et~al.}(2007{\natexlab{b}}){Harris}, {Harris}, {Layden}, \&
  {Wehner}}]{Harris2007}
{Harris}, W.~E., {Harris}, G.~L.~H., {Layden}, A.~C., \& {Wehner}, E.~M.~H.
  2007{\natexlab{b}}, \apj, 666, 903

\bibitem[{{Kalirai} {et~al.}(2006){Kalirai}, {Gilbert}, {Guhathakurta},
  {Majewski}, {Ostheimer}, {Rich}, {Cooper}, {Reitzel}, \&
  {Patterson}}]{Kalirai2006}
{Kalirai}, J.~S., {Gilbert}, K.~M., {Guhathakurta}, P., {et~al.} 2006, \apj,
  648, 389

\bibitem[{{Karachentsev}(2005)}]{Karachentsev2005}
{Karachentsev}, I.~D. 2005, \aj, 129, 178

\bibitem[{{Landolt}(1992)}]{Landolt1992}
{Landolt}, A.~U. 1992, \aj, 104, 340

\bibitem[{{Malin} {et~al.}(1983){Malin}, {Quinn}, \& {Graham}}]{Malin1983}
{Malin}, D.~F., {Quinn}, P.~J., \& {Graham}, J.~A. 1983, \apjl, 272, L5

\bibitem[{{Marleau} {et~al.}(2000){Marleau}, {Graham}, {Liu}, \&
  {Charlot}}]{Marleau2000}
{Marleau}, F.~R., {Graham}, J.~R., {Liu}, M.~C., \& {Charlot}, S. 2000, \aj,
  120, 1779

\bibitem[{{Naab} {et~al.}(2009){Naab}, {Johansson}, \& {Ostriker}}]{Naab2009}
{Naab}, T., {Johansson}, P.~H., \& {Ostriker}, J.~P. 2009, \apjl, 699, L178

\bibitem[{{Oser} {et~al.}(2010){Oser}, {Ostriker}, {Naab}, {Johansson}, \&
  {Burkert}}]{Oser2010}
{Oser}, L., {Ostriker}, J.~P., {Naab}, T., {Johansson}, P.~H., \& {Burkert}, A.
  2010, \apj, 725, 2312

\bibitem[{{Peng} {et~al.}(2004{\natexlab{a}}){Peng}, {Ford}, \&
  {Freeman}}]{Peng2004a}
{Peng}, E.~W., {Ford}, H.~C., \& {Freeman}, K.~C. 2004{\natexlab{a}}, \apjs,
  150, 367

\bibitem[{{Peng} {et~al.}(2004{\natexlab{b}}){Peng}, {Ford}, \&
  {Freeman}}]{Peng2004}
---. 2004{\natexlab{b}}, \apj, 602, 685

\bibitem[{{Peng} {et~al.}(2002){Peng}, {Ford}, {Freeman}, \&
  {White}}]{Peng2002}
{Peng}, E.~W., {Ford}, H.~C., {Freeman}, K.~C., \& {White}, R.~L. 2002, \aj,
  124, 3144

\bibitem[{{Pietrinferni} {et~al.}(2004){Pietrinferni}, {Cassisi}, {Salaris}, \&
  {Castelli}}]{Pietrinferni2004}
{Pietrinferni}, A., {Cassisi}, S., {Salaris}, M., \& {Castelli}, F. 2004, \apj,
  612, 168

\bibitem[{{Pietrinferni} {et~al.}(2006){Pietrinferni}, {Cassisi}, {Salaris}, \&
  {Castelli}}]{Pietrinferni2006}
---. 2006, \apj, 642, 797

\bibitem[{{Rejkuba} {et~al.}(2005){Rejkuba}, {Greggio}, {Harris}, {Harris}, \&
  {Peng}}]{Rejkuba2005}
{Rejkuba}, M., {Greggio}, L., {Harris}, W.~E., {Harris}, G.~L.~H., \& {Peng},
  E.~W. 2005, \apj, 631, 262

\bibitem[{{Rejkuba} {et~al.}(2011){Rejkuba}, {Harris}, {Greggio}, \&
  {Harris}}]{Rejkuba2011}
{Rejkuba}, M., {Harris}, W.~E., {Greggio}, L., \& {Harris}, G.~L.~H. 2011,
  \aap, 526, A123

\bibitem[{{Rejkuba} {et~al.}(2014){Rejkuba}, {Harris}, {Greggio}, {Harris},
  {Jerjen}, \& {Gonzalez}}]{Rejkuba2014}
{Rejkuba}, M., {Harris}, W.~E., {Greggio}, L., {et~al.} 2014, \apjl, 791, L2

\bibitem[{{Rizzi} {et~al.}(2007){Rizzi}, {Tully}, {Makarov}, {Makarova},
  {Dolphin}, {Sakai}, \& {Shaya}}]{Rizzi2007}
{Rizzi}, L., {Tully}, R.~B., {Makarov}, D., {et~al.} 2007, \apj, 661, 815

\bibitem[{{Robin} {et~al.}(2003){Robin}, {Reyl{\'e}}, {Derri{\'e}re}, \&
  {Picaud}}]{Robin2003}
{Robin}, A.~C., {Reyl{\'e}}, C., {Derri{\'e}re}, S., \& {Picaud}, S. 2003,
  \aap, 409, 523

\bibitem[{{Schlegel} {et~al.}(1998){Schlegel}, {Finkbeiner}, \&
  {Davis}}]{Schlegel1998}
{Schlegel}, D.~J., {Finkbeiner}, D.~P., \& {Davis}, M. 1998, \apj, 500, 525

\bibitem[{{Soria} {et~al.}(1996){Soria}, {Mould}, {Watson}, {Gallagher},
  {Ballester}, {Burrows}, {Casertano}, {Clarke}, {Crisp}, {Griffiths},
  {Hester}, {Hoessel}, {Holtzman}, {Scowen}, {Stapelfeldt}, {Trauger}, \&
  {Westphal}}]{Soria1996}
{Soria}, R., {Mould}, J.~R., {Watson}, A.~M., {et~al.} 1996, \apj, 465, 79

\bibitem[{{Stetson}(1987)}]{Stetson1987}
{Stetson}, P.~B. 1987, \pasp, 99, 191

\bibitem[{{Stetson}(1992)}]{Stetson1992}
{Stetson}, P.~B. 1992, in Astronomical Society of the Pacific Conference
  Series, Vol.~25, Astronomical Data Analysis Software and Systems I, ed. D.~M.
  {Worrall}, C.~{Biemesderfer}, \& J.~{Barnes}, 297

\bibitem[{{Stetson}(1994)}]{Stetson1994}
---. 1994, \pasp, 106, 250

\bibitem[{{Stetson}(2000)}]{Stetson2000}
---. 2000, \pasp, 112, 925

\bibitem[{{Williams} {et~al.}(2000){Williams}, {Baum}, {Bergeron}, {Bernstein},
  {Blacker}, {Boyle}, {Brown}, {Carollo}, {Casertano}, {Covarrubias}, {de
  Mello}, {Dickinson}, {Espey}, {Ferguson}, {Fruchter}, {Gardner}, {Gonnella},
  {Hayes}, {Hewett}, {Heyer}, {Hook}, {Irwin}, {Jones}, {Kaiser}, {Levay},
  {Lubenow}, {Lucas}, {Mack}, {MacKenty}, {Madau}, {Makidon}, {Martin},
  {Mazzuca}, {Mutchler}, {Norris}, {Perriello}, {Phillips}, {Postman}, {Royle},
  {Sahu}, {Savaglio}, {Sherwin}, {Smith}, {Stiavelli}, {Suntzeff}, {Teplitz},
  {van der Marel}, {Walker}, {Weymann}, {Wiggs}, {Williger}, {Wilson},
  {Zacharias}, \& {Zurek}}]{Williams2000}
{Williams}, R.~E., {Baum}, S., {Bergeron}, L.~E., {et~al.} 2000, \aj, 120, 2735

\bibitem[{{Woodley} {et~al.}(2010{\natexlab{a}}){Woodley}, {G{\'o}mez},
  {Harris}, {Geisler}, \& {Harris}}]{Woodley2010AJ}
{Woodley}, K.~A., {G{\'o}mez}, M., {Harris}, W.~E., {Geisler}, D., \& {Harris},
  G.~L.~H. 2010{\natexlab{a}}, \aj, 139, 1871

\bibitem[{{Woodley} {et~al.}(2007){Woodley}, {Harris}, {Beasley}, {Peng},
  {Bridges}, {Forbes}, \& {Harris}}]{Woodley2007}
{Woodley}, K.~A., {Harris}, W.~E., {Beasley}, M.~A., {et~al.} 2007, \aj, 134,
  494

\bibitem[{{Woodley} {et~al.}(2010{\natexlab{b}}){Woodley}, {Harris}, {Puzia},
  {G{\'o}mez}, {Harris}, \& {Geisler}}]{Woodley2010ApJ}
{Woodley}, K.~A., {Harris}, W.~E., {Puzia}, T.~H., {et~al.} 2010{\natexlab{b}},
  \apj, 708, 1335

\end{thebibliography}

\end{document}